# Construction of white-light holographic screens

José J. Lunazzi, Daniel S. F. Magalhães

 lunazzi@ifi.unicamp.br, dsouza@ifi.unicamp.br

Laboratório de Óptica, Instituto de Física Gleb Wataghin, P.O.Box 6165, University of Campinas - UNICAMP, 13083-970 Campinas, SP, Brazil

Rolando L. Serra

serra@electrica.cujae.edu.cu

Departamento de Física, Instituto Superior Politécnico "José Antonio Echeverría" (Cujae), Ave. 114, 11901, Marianao, Ciudad de La Habana, CP 19390, Cuba

## Abstract
In this paper we describe one setup employed for the recording of two types of holographic screens that can be used in white-light applications. We show how to obtain holographic screens with areas up to 1370 cm$^2$ and diffraction efficiency of 17%. We analyze the holographic screens in their relevant aspects as to focal lengths, theoretical approach, sizes and diffraction efficiencies specifying when each type is appropriate for particular applications.

## 1. Introduction
The holographic screen [1] is a diffractive optical element (DOE) constructed by holographic techniques in order to obtain the highest directionality such that one projected image can be seen in strictly one direction, while many images can be seen projected simultaneously. It is primarily a holographically-made lens where the vertical and/or horizontal observer's field is extended, allowing for a more comfortable posture of the observer.

The term "holographic screen" is used sometimes in a popular way to designate translucent screens that are used for image projections which produce a phantasmagoric two-dimensional image at the plane of the screen. These screens can be holographically constructed or by other methods. The holographic screens that we will describe here are holographically constructed and are able to produce three-dimensional images.

Most of the white-light applications with the holographic screen are based in a double diffraction process, where the depth of the image is coded by a diffractive element and decoded by the holographic screen [2]. The artist Dan Schweitzer [3] made a simple rainbow Benton hologram to just show to an observer the pure colors it can generate. No three-dimensional object but a plane diffuser was the subject, and color on the whole surface of the hologram changed in a spectral sequence when the observer changed the height of his viewpoint. Steve McGrew made this design popular by producing it embossed [4] and turning it by ninety degrees was the solution found to decode the spectral encoding of depth [5]. This codification-decodification of depth was used to obtain three-dimensional images [6-8] and an application with television purposes was demonstrated [9].

In this work we explain how to obtain holographic screens by holographic techniques that can be used for white-light applications, such as those applications that allow formation of images outside of the plane of the screen [9]. We also analyze different types of installations, optical elements employed in their production reporting a transmission-type holographic screen, with distinct application possibilities than [10], where was achieved the diffraction efficiency of about 17% and maximum size of 1200 cm$^2$.

## 2. White-Light Holographic Screens

Most properties of the holographic screen come from those of a holographic lens. In the recording of a conventional holographic lens the two beams are point sources and their distances to the film characterize the focus of the resulting lens at the wavelength in which the interference pattern is registered [11]. Figure 1 shows the interference of two spherical beams and the distances that characterize the focal length $f$.

$$f = \left( \frac{1}{z_r} - \frac{1}{z_o} \right)^{-1}. \qquad (1)$$

After recording, using quadratic approximations to the spherical waves, the distance $z_i$ from the image point to the holographic lens is given [12] by

$$z_i = \left[ \frac{1}{z_p} + \frac{\lambda_2}{\lambda_1} \left( \pm \frac{1}{z_r} \mp \frac{1}{z_o} \right) \right]^{-1} = \left[ \frac{1}{z_p} \pm \frac{\lambda_2}{\lambda_1} f^{-1} \right]^{-1}, \qquad (2)$$

where the upper set of signs applies for one wave and the lower set for a second wave. These two waves are the result of the amplitude transmittance of the developed transparency. $\lambda_1$ is the wavelength used during the recording process and $\lambda_2$ is the wavelength in the reconstruction. The representation of the distances $z_i$ and $z_p$ are shown in Figure 2, when a monochromatic projector produces a diverging light point that converges as soon as it passes through the holographic lens (HL). This image point becomes the preferred observer's position, where the eye of an observer may receive light coming from the whole surface of the screen, in which images can be produced by using different techniques. The distances $z_i$ from the preferred observer's position to the holographic screen, in the horizontal and vertical types, are given by the same equation 2. Though equation 2 assumes paraxial approximation, this relation continues providing accepted values for angles lesser than 25 degrees between the beams with precision greater than 92%.

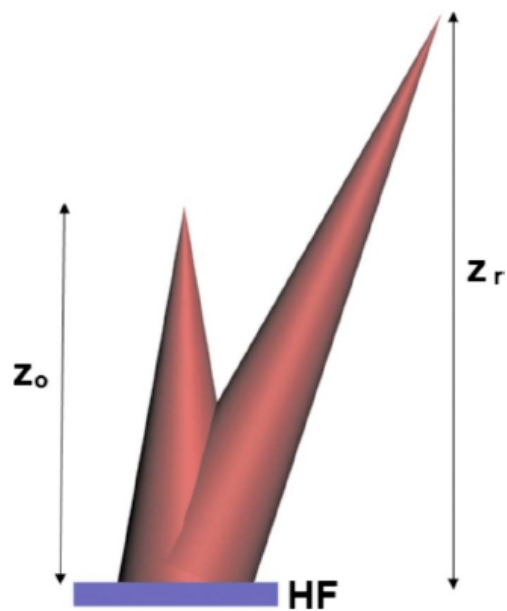

**Fig. 1** Interference of two spherical waves in a photosensitive film (HF) resulting in an off-axis holographic lens. $z_r$ and $z_o$ are the distances from the origin of the reference and object beams to the plane of HF.

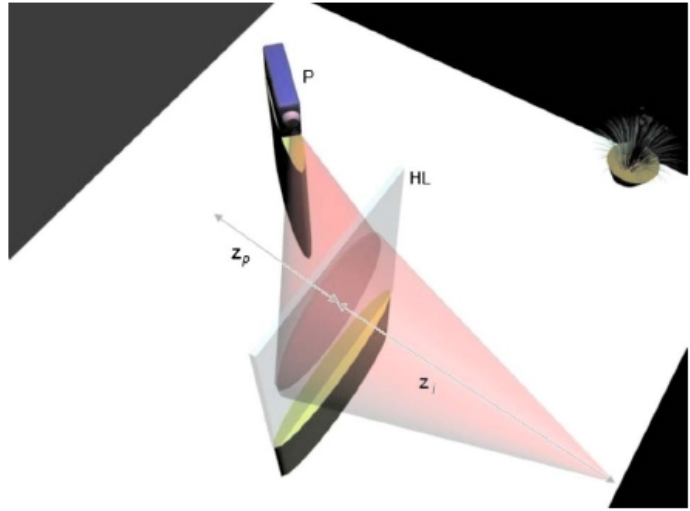

Fig. 2 A monochromatic projector (P) originates a diverging light point that converges as soon as it passes through the holographic lens (HL). $z_p$ is the projection distance, and $z_i$ is the focalization distance, both orthogonal to HL.

For angles greater than 25 degrees we should use a more general equation that can be obtained with geometrical approach. The Figure 3 shows two point sources illuminating a holographic film located at position P and Q. The optical path difference Δl=r+s-t is given by equation 3.

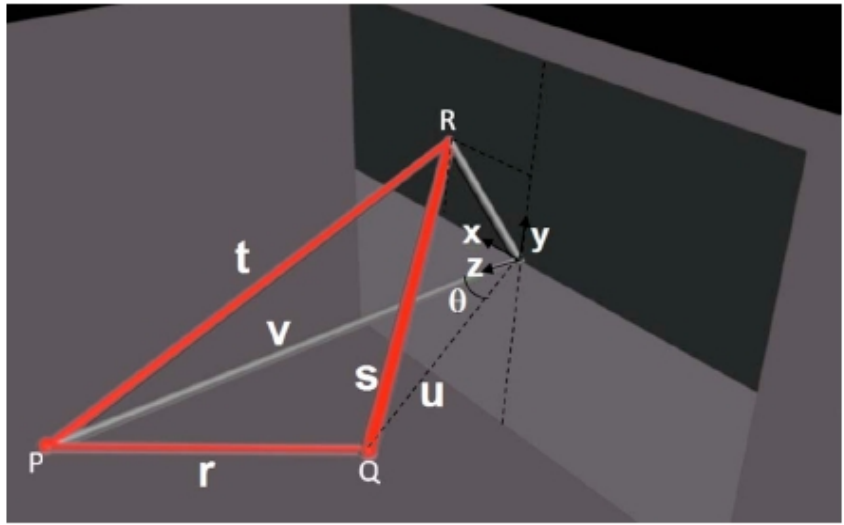

Fig. 3 Interference of two spherical waves.

$$\Delta l = (v^2 + u^2 - 2uv \cos \theta)^{1/2} + [(x_Q - x)^2 + (y_Q - y)^2 + z_Q^2]^{1/2} - (v^2 + x^2 + y^2)^{1/2}, \tag{3}$$

where ($x_Q$, $y_Q$, $z_Q$) are the cartesian coordinates of the point Q (Figure 3). When Δl is a multiple of λ we will have an interference maxima, so the equation 4 will define the level curves when solving (x,y) for each n.

$$(v^2 + u^2 - 2uv \cos \theta)^{1/2} + [(x_Q - x)^2 + (y_Q - y)^2 + z_Q^2]^{1/2} - (v^2 + x^2 + y^2)^{1/2} = n\lambda. \tag{4}$$

To find the focal length of a holographic lens without approximations, we need to determine the spatial frequency in two points of the lens. With these frequencies known we calculate the trajectories of the diffracted rays when the incident ones reach the lens at the angle of the reference beam during recording, these rays are determined using equation 5. In equation 5, $\theta_i$ is the incident angle of the beam reaching the holographic lens in the small part with known spatial frequency ν and $\theta_d$ is diffracted angle of this beam. Finally we discover the convergence point by locating the point where the intersection of the diffracted rays occurs. The focal length *f* is the distance from this point to the lens. The frequencies can be numerically found by the following method:

1. Use the equation 4 with all the values employed at the recording and the point (x=$x_n$, y=0) of the lens and find n. $x_n$ is the x coordinate of the first point for which we will discover the frequency.

2. Calculate x=$x_{n+1}$ to the known (n+1) diffraction fringe.

3. Calculate the frequency ν by equation 6.

4. Repeat this process to find a ν of a different $x_n$.

$$\sin \theta_i - \sin \theta_d = m\lambda\nu, \tag{5}$$

$$\nu = |x_{n+1} - x_n|^{-1}. \tag{6}$$

In the white-light holographic screens, the observer's field is vertically extended by the use of a diffuser during the holographic recording, for practicality. We use the label "Horizontally- Dispersive white- light Holographic Screen" (HDHS) for those screens in which the diffractive dispersion occurs horizontally and the diffuser's image is vertical [9] and we use the label "Vertically- Dispersive white- light Holographic Screen" (VDHS) for those screens in which the diffractive dispersion occurs vertically. These two different holographic screens can be obtained with a simple change in the holographic setup.

The result presented in equation 4 can be extended to find the level curves of a HDHS or a VDHS, assuming each point of the diffuser mutually coherent with the reference beam. First we construct the function that describes the pattern of fringes by drawing a level curve for each n of equation 4, this represents the structure created by one point of the diffuser of the Figure 6. Displacing the coordinates of the point Q in Figure 3 of a small amount we get another group of curves. Numerical mapping the (x,y) coordinates of the superimposed curves, we have the total structure created by the fringes of all the diffuser points. This function can be used to obtain the transmittance function [13] of the HDHS or the VDHS, depending of the choices of the displacements of the point Q.

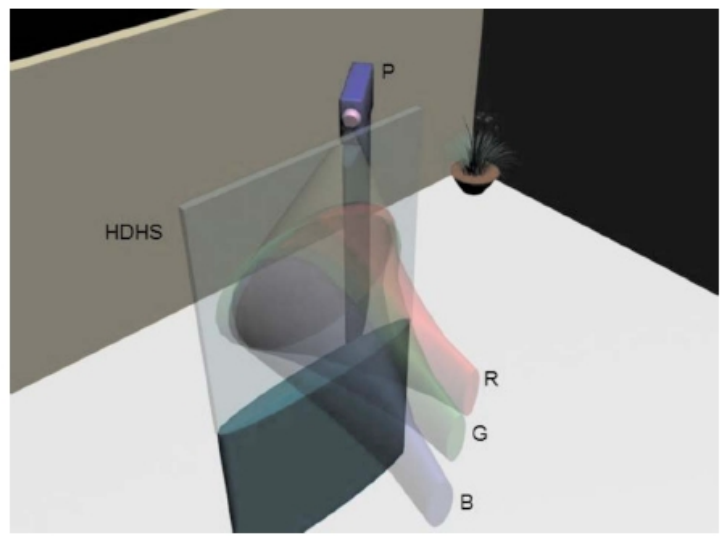

Fig. 4 The horizontally dispersive white- light holographic screen (HDHS) being illuminated by a diverging white-light beam. R, G, and B are the representations of the red, green, and blue images of the diffuser. (Color online only).

When the HDHS is illuminated by a diverging white-light beam, each wavelength will converge at a different position because it constructs the diffuser's images, as shown in Figure 4. When a diverging white-light beam illuminates a VDHS, an overlapping of the diffuser's image takes place as shown in Figure 5, which may be conveniently designed to obtain a full color image. A HDHS is employed by laterally projecting a sequence of views being encoded in a spectral sequence of wavelengths. The observer sees this sequence as a horizontal parallax, which may be continuous if the sequence is generated through a diffraction process [1, 6-8]. A VDHS bases its horizontal parallax on the geometrical position of the projectors, up or down the screen. The parallax may be made continuous by employing a continuous projecting source, as the shadows from a linear horizontal source [14] but one may find a more practical application by using two projectors to make a stereoscopic setup [15].

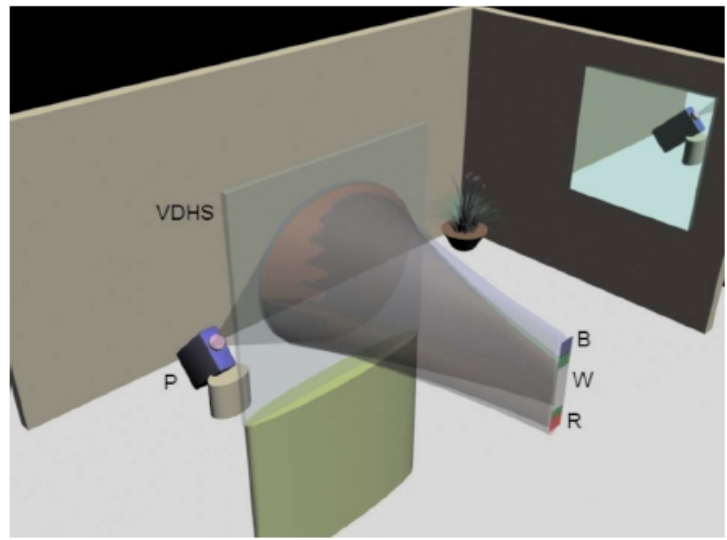

**Fig. 5** The vertically dispersive white-light holographic screen (VDHS) being illuminated by a diverging white-light beam. The projector P illuminates the VDHS, and an overlapping of the diffuser's image appears in W. At the edges, the blue (B) and red (R) images are shown. (Color online only).

## 3. Holographic Recording of White-Light Holographic Screens

Above we describe the holographic recording method to obtain the two different kinds of white-light holographic screens. The used holographic film in these recordings was the Agfa Gevaert 8E75 HD [16] and a stabilized holographic table was employed.

For most applications with white-light is convenient to extend the vertical observer's field [9], the idea of one diverging point [12] can be extended to an array of diverging points in the origin of the object beam. This array could be a diffusing vertical straight line, where each diffusing point of the line illuminates the entire film. The width of this line must be as thin as possible to avoid the overlapping of wavelengths to get the best directionality.

The Figure 6 shows the design of the experimental setup to obtain the horizontally-dispersive white-light holographic screens (HDHS). The collimated beam of an He-Ne laser with power 30 mW (L) is divided by a variable beam splitter (BS), subsequently each beam is reflected by a plane mirror (M). The reference beam reaches the spatial filter (SP) and illuminates the holographic film (HF). The object beam reaches a cylindrical lens (CL) of focal length 3.8 mm and it makes the beam to diverge creating a diffusion line at the diffuser (D). The diffuser line had 2 mm wide and length of 200 mm. Each diffuser point of the diffuser line illuminates the film. The material that causes the diffusion was a triple layer of plastic sandwiched by two thin glass plates. The employed angle between the rays was 45°.

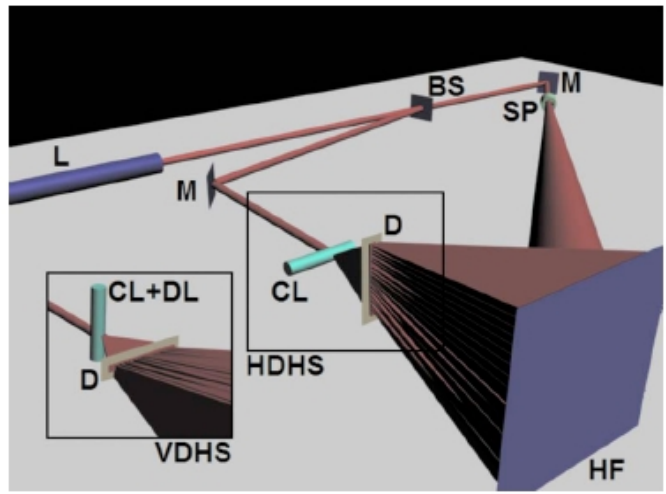

Fig. 6 Experimental setup utilized in the holographic recordings of the horizontally dispersive white-light holographic screen (HDHS) and the vertically dispersive white-light holographic screen (VDHS).

To obtain vertically-dispersive white-light holographic screens (VDHS) we just need to rotate the cylindrical lens (CL) by 90°, as can be seen at the left of Figure 6, and introduce a simple diverging lens (DL) closed to CL. The resulting diffuser line had 6 mm wide and length of 200 mm, this small difference in the width of the diffuser line increase the observer's field.

The distances employed in the recording of both holographic screens are presented in Table 1.

Table 1 Distances employed in the recording of the holographic screens.

| Type of screen | D-HF (cm) | CL-D (cm) | SP-HF (cm) |
|---|---|---|---|
| HDHS | 70 ± 1 | 20 ± 1 | 237 ± 3 |
| VDHS | 58.5 ± 0.5 | 29 ± 1 | 237 ± 3 |

The vertically-dispersive white-light holographic screen (VDHS) must be illuminated from above or from below, which will cause a vertical dispersion of the diffracted light, as can be seen in Figure 5. An application of this screen is a goggle-less stereoscope that could be used in medical applications [15].

## 4. Experimental Results

Using the experimental setups described we obtained HDHS and VDHS. The exposure time was about 40 seconds with the Spectra-Physics He-Ne laser of 30mW. After recording we employed the chemical development described in Table 2. The ideal optical density to the Agfa film before the development should be about 2.5 and the time of development is 2 min. Subsequently we utilized two successive bleaching processes as shown in table 2 [17]. Both processes spend about 2 minutes when the optical density is 2.5.

Table 2 Chemical development and bleaching used with Agfa-Gevaert 8E75 HD.

| Solution A | Solution B | 1st bleach | 2nd bleach |
|---|---|---|---|
| Pyrogallol—10 g | $Na_2CO_3$—60 g | $(NH_4)_2Cr_2O_7$—3 g | $KMnO_4$—0.8 g |
|  |  | $H_2SO_4$—1.5 ml | $H_2SO_4$—10 ml |
| Distilled water—1 l | Distilled water—1 l | Distilled water—1 l | Distilled water—1 l |

The selection of the angle of θ=45° occurs to avoid the overlapping of the zero diffraction order with the first one, which would disturb the observer's view. Using equation 4 and the method described in section 2, we determine the theoretic focal lengths of the screens. We measured them by projecting a collimated laser beam with the same wavelength that in the recording with θ=45° and we measured the z coordinate of the focalized image point. The values for the HDHS and for VDHS are presented in Table 3.

Table 3 Focal length of the holographic screens.

| Type of screen | Measured<br>f to λ = 632 nm | Theoretical<br>f to λ = 632 nm |
|---|---|---|
| HDHS | 85 ± 2 cm | 82.2 cm |
| VDHS | 65 ± 2 cm | 66.8 cm |

The use of the diffuser in the HDHS increases the vertical observation field to 8 cm when the observer is at 2 m in front of the screen in white-light applications. At the same recording conditions except by the diffuser, a conventional holographic lens offers no more than 1 cm of vertical observation field because the observer's field is correlated with the size of the aperture of the projecting apparatus. The obtained diffraction efficiency was 17±1%. Though the 45 degrees angle, which could mean a non-negligible Bragg effect and causes a strong dependence of the diffraction efficiency with the wavelength, we measured the diffraction efficiencies for two different wavelengths $\lambda$=632 nm and $\lambda$=520 nm) and the difference between then is fewer than 2%. The resulting Q-factor for the setup [18] was about 19, this explain the small difference of efficiencies, i.e., the screens have many of the properties of a thin hologram. Due to this small percentual difference we can consider the diffraction efficiency of the white-light holographic screens independent of the wavelength for screens with $\theta<45°$.

Furthermore, the size of 1370 cm$^2$ of both screens allows the projection of large images. The focal lengths of the screens allow the developing of image devices with suitable size [9].

The width of the 8E75HD Agfa film of about 1 meter limits the size of the screens, but in our case the limitation on size is intrinsically linked to the limitation on efficiency due to the wavelength stability of the used He-Ne laser. The bigger the size of the screen, the greater the optical path difference in the interference of the object and reference beams, reducing the diffraction efficiencies because of the movement of the fringes during exposure. One of the authors performed a 1.15 m x 0.80 m screen by using a pulsed laser [19] employed for the enlargement of the holograms.

## 5. Conclusion

In this work we discussed the setup used in the recording of two different types of holographic screens looking forward their applications with white light as previously published. We showed how to obtain holographic screens of 1370 cm$^2$ with diffraction efficiency of 17%. We analyzed the holographic screens in the relevant aspects as their focal lengths, sizes and diffraction efficiencies specifying when each type is appropriate for some kind of developed application.

## Acknowledgements

The autors thanks A. C. Costa for allowing the use of some dependeces of his laboratory, the family Baumstein for donation of special optical elements and to John R. Smith of the University of California - Davis for the english revision.

## Biography

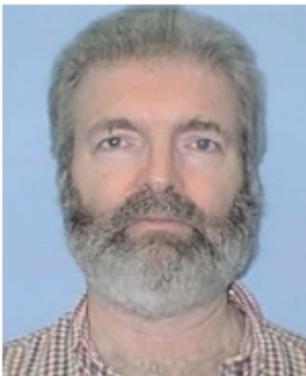

José Joaquín Lunazzi was born In La Plata, Argentina, In 1948. He received his M.Sc. degree in 1970 and his Ph.D. degree in 1975, both in physics, from La Plata National University. He constructed his first hologram in 1969 as a consequence of his interest in stereo photography, which he performed at the age of 14. He  joined Campinas State University in 1976, working in optical metrology, interferometry and holography. Dr. Lunazzi introduced holographic techniques in Latin America and has taught optics at every level, from postgraduate courses to courses for teachers at secondary and primary levels. He has also taught the use of simple materials in general courses  for photographers, university students, artists, and the general public.

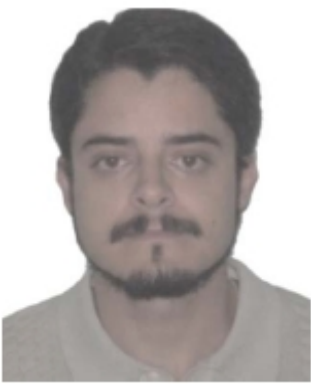

Daniel S. F. Magalhães received a B.S. degree in Physics (2003) and M.S. degree (2005) in Diffractive Optics and Imaging from Universidade Estadual de

Campinas, Campinas, Brazil. He received his Ph.D. degree (2009) from the same university in the subject of Construction of Holographic Screens and their Applications.

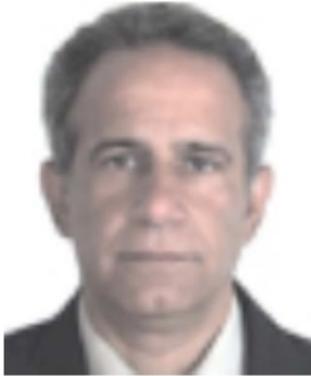

Rolando Serra Toledo received the B.S. degree in Physics (1981) from Instituto Superior Pedagógico Enrique José Varona, M.S. degree in Optics and Lasers (1990) and Ph.D. degree in Pedagogical Sciences (1990), both from Instituto Superior Politécnico José A. Echeverría, Cuba.

He has conducted trainings in Holography and laser applications in Hungary and Spain. He has performed graduate courses in holography and non-destructive optical testing in Cuba and Spain. He has about 30 international published articles, 3 registered patents in Cuba and has participated in several international projects.

He is currently a professor at Instituto Superior Politécnico José A. Echeverría and the director of the Holographic Research Group in Cuba, which is considered the national leader in the holographic thematic.